RESEARCH ARTICLE

# Sensitivity analyses for effect modifiers not observed in the target population when generalizing treatment effects from a randomized controlled trial: Assumptions, models, effect scales, data scenarios, and implementation details

Trang Quynh Nguyen[1,2]*, Benjamin Ackerman[2], Ian Schmid[1], Stephen R. Cole[3], Elizabeth A. Stuart[1,2,4]

1 Department of Mental Health, Johns Hopkins Bloomberg School of Public Health, Baltimore, MD, United States of America, 2 Department of Biostatistics, Johns Hopkins Bloomberg School of Public Health, Baltimore, MD, United States of America, 3 Department of Epidemiology, Gillings School of Global Public Health, University of North Carolina, Chapel Hill, NC, United States of America, 4 Department of Health Policy and Management, Johns Hopkins Bloomberg School of Public Health, Baltimore, MD, United States of America

* tnguye28@jhu.edu

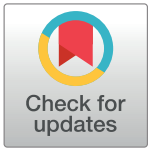

**Data Availability Statement:** All relevant data are within the manuscript and its Supporting Information files.

**Funding:** This work was supported in part by NSF grant DRL-1335843 (co-PIs E. A. Stuart and R. B. Olsen), PCORI grant ME-150227794 (co-PIs I. Dahabreh and E. A. Stuart), NIDA grant R01-DA036520 (PI R. Mojtabai), and NIAID grant R01AI100654 (PI S. R. Cole). The funders had no

## Abstract

### Background

Randomized controlled trials are often used to inform policy and practice for broad populations. The average treatment effect (ATE) for a target population, however, may be different from the ATE observed in a trial if there are effect modifiers whose distribution in the target population is different that from in the trial. Methods exist to use trial data to estimate the target population ATE, provided the distributions of treatment effect modifiers are observed in both the trial and target population—an assumption that may not hold in practice.

### Methods

The proposed sensitivity analyses address the situation where a treatment effect modifier is observed in the trial but not the target population. These methods are based on an outcome model or the combination of such a model and weighting adjustment for observed differences between the trial sample and target population. They accommodate several types of outcome models: linear models (including single time outcome and pre- and post-treatment outcomes) for additive effects, and models with log or logit link for multiplicative effects. We clarify the methods' assumptions and provide detailed implementation instructions.

### Illustration

We illustrate the methods using an example generalizing the effects of an HIV treatment regimen from a randomized trial to a relevant target population.

## Conclusion

These methods allow researchers and decision-makers to have more appropriate confidence when drawing conclusions about target population effects.

## Introduction

Randomized controlled trials (*trials*) are often used to inform policy and practice for broad populations. A well designed and implemented trial allows consistent estimation of the average treatment/intervention effect (ATE) for the trial sample. We refer to this as the *Study-specific ATE* (SATE). If the question is whether that treatment should be used for people in a certain population (*target population*), then of interest is the *Target population ATE* (TATE). Since trial samples are often not representative of target populations, SATE may not be a good estimate of TATE. SATE departs from TATE if the trial sample differs from the target population with respect to the distribution of treatment effect modifiers.

Methods exist to use trial data to estimate TATE, e.g., [1–3], assuming treatment effect variation is explained by pre-treatment variables that are observed in both the trial and target population. Often, however, the variables observed for the target population are limited compared to those measured in the trial [4, 5]. This paper presents simple methods to assess the sensitivity of TATE estimates to effect modifiers observed in the trial but not in the target population. The paper starts from the simple case of additive effects based on an uncomplicated linear model (previously addressed in [6]) and extends to cases with more complex models and multiplicative effects. We clarify the assumptions of these methods for a general audience and provide detailed implementation instructions.

We illustrate the sensitivity analyses using an example based on the AIDS Clinical Trial Group (ACTG) 320 Study [7]. This trial randomized HIV-infected adults to two antiretroviral regimens: (i) two nucleoside reverse-transcriptase inhibitors (AZT or d4T and 3TC) plus a protease inhibitor (Indinavir), and (ii) AZT/d4T and 3TC only—referred to as new and old treatment, respectively. The trial found that relative to the old treatment, the new treatment lowered the hazard of AIDS and/or death. Cole & Stuart [1] generalized this effect to the population of people diagnosed with HIV in the US in 2006, using a set of covariates observed in both the trial and target population. Our example is based on this trial-population pair. We consider a different outcome, CD4 count (number of T-CD4 cells per ml blood), which is an important indicator of immunity status. We use the same target population dataset that Cole & Stuart created for [1] based on the CDC-estimated joint distribution of demographic characteristics in this population [8]. As the trial data are not for open access, for this illustration, we use a synthetic trial dataset created to mimic the distributions in the real trial data.

## Methods for the simplest case: Additive effects on potential outcome based on a linear causal model

For this case, the basic mathematical results of the methods and one of their key assumptions were developed in [6]. The current paper elaborates on the full set of assumptions required by these methods, and pays close attention to details relevant to their effective use, such as different data scenarios and corresponding implementation instructions, including variation in weighting procedures for the method that involves weighting.

### Trial sample, target population data, effect definitions

Consider a trial evaluating the effect of treatment $A$ on outcome $Y$. Let $S = 1$ denote trial participation, $P = 1$ denote membership in the target population. In Fig 1, the trial sample is drawn from the target population (i.e., individuals $i$ with $S_i = 1$ also have $P_i = 1$); and the purpose is to generalize trial results back to the target population. If the trial sample is drawn from a different population (i.e., $S = 1$ is outside of $P = 1$), the problem is to transport [9] results to the target population.

We present the methods using a binary treatment. Let $Y_i(a)$ denote individual $i$'s potential outcome [10] if treatment were set to $a$, with $a = 1, 0$ (treatment or control). Treatment effect for individual $i$ is defined on the additive scale as $Y_i(1) - Y_i(0)$, the difference between potential outcomes under treatment and under control.

SATE is defined as the average treatment effect in the trial sample, $E[Y(1) - Y(0)|S = 1]$. Due to randomization, SATE is unbiasedly estimated, for example, by the difference in mean outcome between the trial's treatment arms. Of interest, however, is TATE, defined as $E[Y(1) - Y(0)|P = 1]$, the average treatment effect in the target population.

The sensitivity analyses require data on the distribution of pre-treatment covariates in the target population. Such data may come from a full population ($P = 1$) dataset, a representative ($S = 2$) sub-sample, or population summary statistics (see Fig 1).

### Assumptions

Several assumptions are required. The first assumption is that the trial has internal validity (A1), which itself consists of several conditions listed in Table 1. We will not discuss interval validity further, but this is a key assumption. On top of this, we need a set of assumptions commonly used in generalization [1–3]: across-setting treatment variance irrelevance (A2), trial coverage of target population ranges of the effect modifiers (A3), conditional ignorability for treatment effects (A4), and consistent measurement and no measurement error (A5)—see detailed explanations of and practical comments on these assumptions in Table 1.

Assumptions A1-A5 are sufficient for estimating TATE if all effect modifiers are observed in both the trial and target population, via either G-computation or weighting the trial sample to the target population [1–3, 11, 12]. When some effect modifiers are not observed in the target population, however, such strategies fail. In this case, in order to use the proposed sensitivity analyses to glean some information on TATE, we make the additional assumption (A6) of a

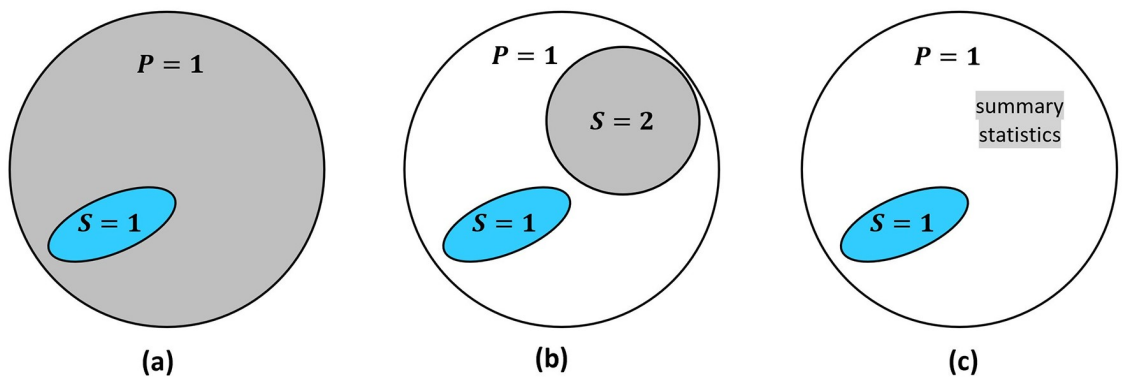

**Fig 1. Several data source scenarios for generalization from the trial ($S = 1$) to the target population ($P = 1$).** (a) the trial sample and a full population dataset; (b) the trial sample and a dataset ($S = 2$) that is representative of the population; (c) the trial sample and some summary statistics about the population.

**Table 1. Key assumptions.**

| Assumption | | Details |
|---|---|---|
| A1. | Internal validity of the trial | We make all the assumptions required for the trial's internal validity, e.g., conditional ignorability of treatment assignment, positivity, treatment variance irrelevance, no interference, etc. |
| A2. | Across-setting treatment variation irrelevance | When a treatment is applied to the target population, it is administered in settings that are likely different from the trial setting. The assumption is that the differences in the treatment that result do not change its effect.<br>This assumption is non-trivial in some settings and should be examined explicitly in generalization practice. In this paper, we simply make the assumption; its plausibility is not the problem we currently aim to discuss. |
| A3. | Treatment effect modifiers coverage | Treatment effects depend on a set of pre-treatment variables observed in the trial (denoted $Z$ if observed and $V$ if not observed in the target population), and the range of these (observed and unobserved) effect modifiers in the target population is covered by their range in the trial. The rationale is that if there are members of the target population with effect modifier values outside the range observed in the trial, we cannot use trial data to learn about their treatment effects. This assumption, which is similar to positivity, is formally $\Pr(S = 1 \mid Z = z, V = v) > 0$ if $\Pr(Z = z, V = v \mid P = 1) > 0$ where $\Pr[\cdot]$ is the probability function [17]. This assumption can be checked for $Z$ variables but not for $V$. |
| A4. | Conditional sample ignorability for treatment effects [3] | For an individual (in the trial or the target population) with effect modifiers within the target population range, "sample membership" (i.e., whether the individual is in the trial or in the target population) does not carry any information about the treatment effect once we condition on the effect modifiers. For additive effects, this is formally $[Y(1) - Y(0)] \perp\!\!\!\perp \{S, P\} \mid Z, V, (S = 1 \text{ or } P = 1)$.<br>This assumption allows generalizing treatment effects from individuals in the trial to individuals in the target population with similar patterns of the effect modifiers.<br>This assumption is less strict than conditional sample ignorability for potential outcomes, $\{Y(1), Y(0)\} \perp\!\!\!\perp \{S, P\} \mid X, Z, V, (S = 1 \text{ or } P = 1)$ where $X$ denotes predictors of the potential outcomes that are not part of the set of effect modifiers $\{Z, V\}$. Yet this assumption is tied to the scale on which effects are defined. |
| A5. | Consistent measurement and no measurement error | The trial's internal validity requires no systematic measurement error of a continuous outcome and no misclassification of a categorical outcome. Here we also need to assume that all covariate measurements are without error, and that $X$ and $Z$ are measured the same way between the trial and the target population. |
| A6. | A specific causal model | We assume a causal model with effect modification; effects should be defined on a scale that connects naturally to the model.<br>The main methods section assumes a linear causal model (with interaction terms) and defines effects on the additive scale. |

causal model. In this section we assume the linear causal model

$$E[Y_i(a)] = \beta_0 + \beta_a a + \beta_x X_i + \beta_z Z_i + \beta_{za} Z_i a + \beta_v V_i + \beta_{va} V_i a, \qquad \text{(M1)}$$

where potential outcomes are influenced by treatment condition $a$ and baseline covariates $X, Z, V$, all observed in the trial. $X, Z$ and $V$ may be multivariate; the use of univariate notation is to simplify presentation. $Z$ and $V$ both denote effect modifiers; the difference is $Z$ is observed in the target population while $V$ is not.

(The letter M in the equation label M1 indicates that this is a causal <u>m</u>odel. We will also use T in labels to indicate <u>T</u>ATE formulas, and R to indicate <u>r</u>egression models).

This model assumption should not be made lightly. With a binary outcome, for instance, this model would imply additive effects on the risk difference scale, which may be inappropriate.

(See the next section for an extension of these methods to a broader class of models). Also, the selection of which pre-treatment variables to include in the model and which variables to consider effect modifiers requires serious investigation, and is discussed is great length in relevant literature, e.g., [13–16]. With the current focus on sensitivity analysis, we presume that such a model has been selected. If unsure whether a variable is an effect modifier (e.g., because its interaction with treatment has a non-negligible coefficient but its p-value is large), our recommendation is to treat it as one for the purpose of TATE estimation, since trials usually lack power to investigate effect modification.

## TATE formula

Based on the causal model M1, individual $i$'s treatment effect has expectation $E[Y_i(1) - Y_i(0)]$ = $\beta_a + \beta_{za} Z_i + \beta_{va} V_i$, therefore

$$\text{TATE} = \beta_a + \beta_{za}E[Z|P = 1] + \beta_{va}E[V|P = 1], \tag{T1}$$

where $E[Z|P = 1]$ and $E[V|P = 1]$ are the means of $Z$ and $V$ in the target population. Under Assumptions A1-A6, $\beta_a$, $\beta_{za}$ and $\beta_{va}$ are the same for the trial sample and target population, and can be estimated using trial data.

The problematic quantity in this formula is $E[V|P = 1]$, because we do not observe $V$ in the target population. However, if we specify a plausible range for $E[V|P = 1]$ (the *sensitivity parameter*), we obtain a range of TATE estimates. We outline here two ways to do this, and provide detailed implementation instructions in Table 2.

### Method 1: Outcome-model-based sensitivity analysis

This method requires an estimate for $E[Z|P = 1]$ (target population mean $Z$), but not a target population dataset. It involves: (1) obtaining an estimate for $E[Z|P = 1]$; (2) specifying a plausible range for the sensitivity parameter $E[V|P = 1]$; (3) fitting to trial data the regression model

$$E[Y|A, X, Z, V] = \beta_0 + \beta_a A + \beta_x X + \beta_{za}ZA + \beta_v V + \beta_{va}VA; \tag{R1}$$

(4) combining the estimated $E[Z|P = 1]$ and specified $E[V|P = 1]$ with model coefficients to obtain TATE estimates (using formula T1); and (5) plotting results against the sensitivity parameter.

### Method 2: Weighted-outcome-model-based sensitivity analysis

If a target population dataset is available, an alternative is to weight the trial sample to mimic the target population distribution of $X, Z$, before implementing the same steps as in method 1.

Simulations (see description and full results in the S1 Appendix) found that Method 2 has an advantage over method 1 with respect to bias; it provides some protection against bias due to misspecification of the outcome model. Specifically, if the outcome model is misspecified with respect to $Z$, method 1 is biased, but method 2 is unbiased because the weighting adjusts for the difference between the trial and target population in effect variation due to the difference in distribution of $Z$. Also, if the outcome model is misspecified with respect to $V$ then both methods are biased, but if $V$ is positively correlated with $Z$ in the trial and influences treatment effect in the same direction as $Z$, method 2 is less biased than method 1, because the weighting adjustment for $Z$ helps partially adjust for $V$. Method 2's disadvantage is that it has larger variance than method 1, and suffers from some degree of variance underestimation— the variance estimated based on the fitted outcome model is on average smaller than true variance. Bias and variance combined leads to the two methods both performing well when the outcome model is correctly specified, with the method 1 confidence interval (CI) having nominal coverage (about 95%) and the method 2 CI's coverage being only slightly smaller (around 93-94%). When the outcome model is misspecified, due to bias reduction, method 2's CI generally has better variance than method 1's.

**Table 2. Implementation instructions.**

| | **Method 1: Outcome-model-based sensitivity analysis** |
|---|---|
| Step 1 | Obtain an estimate for $E[Z\|P = 1]$ (mean of $Z$ in the target population), with confidence limits to reflect uncertainty (unless it is known with certainty, e.g., from a full population dataset). |
| Step 2 | Specify a plausible range for $E[V\|P = 1]$ (aka the *sensitivity parameter*).<br>• This range should ideally be informed by knowledge about this variable from other data or from the literature regarding the target population or similar populations.<br>• When little information is available, a wide range can be used so that consumers of the research could be selective in interpreting the results based on information they may have on this parameter. |
| Step 3 | Fit to the trial data the regression model R1. |
| Step 4 | For each of the lower and upper ends of the range specified for $E[V\|P = 1]$, obtain a corresponding estimate of TATE (including point estimate and confidence limits).<br>Suppose that for $E[Z\|P = 1]$, we have a point estimate of 2 and 95% confidence interval of (1.5, 2.5); and for $E[V\|P = 1]$, we specify a plausible range of 30 to 70. TATE corresponding to one end of this range, e.g., the lower end ($E[V\|P = 1] = 30$), is estimated as follows:<br>• Point estimate: Take a linear combination of the coefficients from model R1—based on the TATE formula T1—using the point estimate 2 of $E[Z\|P = 1]$, that is, $(\beta_a + 2\beta_{za} + 30\beta_{va})$. This can be done using the lincom statement after fitting the model in Stata or using the estimate statement when specifying the model in SAS. The output for this linear combination includes a point estimate, standard error and confidence interval. Take the point estimate of this linear combination as the point estimate for TATE.<br>• Confidence limits: Use the confidence limits (1.5 and 2.5) of $E[Z\|P = 1]$ to take two additional linear combinations: $(\beta_a + 1.5\beta_{za} + 30\beta_{va})$ and $(\beta_a + 2.5\beta_{za} + 30\beta_{va})$. Consider the confidence limits of these linear combinations: take the more extreme of their two upper confidence limits, and the more extreme of their two lower confidence limits, as the confidence limits for TATE. |
| Step 5 | Plot the range of TATE with confidence bounds (y-axis) against the range specified for the sensitivity parameter $E[V\|P = 1]$ (x-axis). Specifically,<br>• Plot the TATE estimates corresponding to the two ends of the range obtained in step 4, each with three points, one for the point estimate and two for the confidence limits; and<br>• Connect the two point estimates, the two lower confidence limits, and the two upper confidence limits, using three straight lines. |
| | **Method 2: Weighted-utcome-model-based sensitivity analysis** |
| Step 0 | Weight the trial sample so that it resembles the target population with respect to $X$, $Z$. |
| Steps 1-2 | Same as in method 1 |
| Step 3 | Fit model R1 to the weighted trial sample. |
| Step 4-5 | Same as in method 1 |

Given these findings, we generally recommend method 2 (with the weighting), unless the two methods agree on TATE point estimates, in which case method 1's unweighted results can be used.

**Weighting procedures.** Method 2 involves weighting the trial sample to mimic the target population distribution of $X$, $Z$. In most situations where a population dataset (either $P = 1$ or $S = 2$—see Fig 1) is available, we use *weighting by the odds* [3, 18]. The exception is when trial participants are part of AND can be identified within the population dataset, then *inverse probability weighting* is used. If only population summary statistics are available, weighting is generally not used. However, if information on the joint distribution of $X$, $Z$ in the target population is available, and $X$, $Z$ are discrete with few combined categories, weighting may be implemented. See Table 3 for details on weights computation in these cases. For why they apply, see the S2 Appendix.

**Table 3. Weighting procedures for different target population data scenarios.**

| Target population data | Weighting procedures |
|---|---|
| A $P = 1$ dataset is available. Trial participants cannot be identified in this $P = 1$ dataset. | ***Weighting-by-the-odds***, i.e., weight the trial participants using weights computed as follows |
| A $S = 2$ dataset is available. Trial participants are either not part of the $S = 2$ sample, or if they are, they cannot be identified in this $S = 2$ dataset. | 1. stacking the trial and target population datasets into one dataset, and creating a new variable $S'$ coded $S' = 1$ for observations from the trial dataset, and $S' = 0$ for observations from the target population dataset |
|  | 2. fitting a model using $X$ and $Z$ to predict $S'$; and for each trial participant, obtaining the predicted probability of $S' = 1$ from that model (aka the trial *participation score*), $ps_i = \Pr(S' = 1 \mid X_i, Z_i)$ |
|  | 3. for every trial participant, computing the weights as $W_i = (1 - ps_i)/ps_i$, the odds of being in the target population dataset. |
| A $P = 1$ dataset is available. Trial participants are identified in this $P = 1$ dataset. | ***Inverse-probability-weighting***, i.e., weight the trial participants using weights computed as follows |
| A $S = 2$ dataset is available. Trial participants are part of the $S = 2$ sample, and are identified in this $S = 2$ dataset. | 1. using the target poplation dataset, creating a new variable $S'$ coded $S' = 1$ for observations that belong to the trial participants, and $S' = 0$ for the remaining observations |
|  | 2. fitting a model using $X$ and $Z$ to predict $S'$; and for each trial participant, obtaining the predicted probability of $S' = 1$ from that model (aka the trial *participation score*), $ps_i = \Pr(S' = 1 \mid X_i, Z_i)$ |
|  | 3. for every trial participant, computing the weights as $W_i = 1/ps_i$, the inverse of the probability of participating in the trial. |
| Information on the joint distribution of $\{X, Z\}$ in the target population is available. $X, Z$ are categorical with a small number of combined categories. | ***Ratio-of-probability-weighting***, specifically, weight the trial participants using weights computed by the formula $W_i = \dfrac{\Pr(X = X_i, Z = Z_i \mid P = 1)}{\Pr(X = X_i, Z = Z_i \mid S = 1)}$, where the numerator and denominator are the prevalences of the $\{X_i, Z_i\}$ pattern in the target population and in the trial sample, respectively. |

An additional note: The desired result of weighting is that the arms of the trial (i) each mimics the target population in the distribution of $X, Z$, and (ii) remain similar to each other in the distribution of $X, Z, V$. We recommend a two step procedure: first checking balance between the trial arms and adjusting via within-trial reweighting if needed; and then weighting the (adjusted) trial sample to the target population. We do not recommend weighting each trial arm to the target population separately, because it may distort between-arms balance on variables not observed in the target population.

## What if an effect modifier is not even observed in the trial?

There are times when instead of an effect modifier observed in the trial but not in the target population, researchers are concerned about effect modifiers that were not measured in the trial. This may be a specific variable, e.g., addiction severity was not measured in a substance abuse treatment trial, but it is suspected to modify treatment effect and it may very well be distributed differently between the trial and the target population. Or it may be generic, when researchers are concerned that there is effect modification by unknown factors.

The question is whether the above-described sensitivity analyses can be extended to cover an effect modifier $U$ (be it a specific or generic variable) that is unobserved in the trial. Unfortunately, the answer is no. It is clear from the causal model

$$\mathrm{E}[Y_i(a)] = \beta_0 + \beta_a a + \beta_x X_i + \beta_z Z_i + \beta_{za} Z_i a + \beta_u U_i + \beta_{ua} U_i a$$

that if we use the approach without weighting, we would need an estimate of $\beta_{za}$, which is unidentified from the trial because $U$ is not observed. If we use the weighting approach and manage to equate the mean of $Z$ between the trial sample and the target population, then we could do without $\beta_{za}$ but instead would have to deal with the mean of $U$ in the weighted trial sample, an obscure quantity that is not suitable to serve as a sensitivity parameter. (For technical details, see the S3 Appendix).

Note that this is a correction of the $U$ case results reported in [6]; the appendix explains the error in those previous results.

## Method extensions

Turning our attention back to effect modifiers that are observed in the trial but not in the target population ($V$), note that the last section (and [6]) addressed a simple setting. We now offer two extensions of the sensitivity analyses to more complex situations.

### Extension 1: When both pre- and post-treatment outcomes are available and are modeled using random intercepts models

When both pre- and post-treatment measures of the outcome are available, there are several options for modeling such data. One option is to treat the pre-treatment outcome measure as a baseline covariate. Another option, which we consider here, is to model the combination of both pre- and post-treatment outcomes using random intercepts models.

The simplest random intercepts model in this case is the model without covariates

$$\mathrm{E}[Y_{ij}|A_i, F_{ij}] = c_{0i} + \gamma_0 + \gamma_a A_i + \gamma_f F_{ij} + \gamma_{fa} F_{ij} A_i,$$

where $i$ indexes person, $j$ indexes observation (each person has two observations), $A$ indicates treatment arm, $F$ indicates whether the observation is pre-treatment ($F_{i1} = 0$) or post-treatment ($F_{i2} = 1$), and the coefficient $\gamma_{fa}$ of $FA$ represents treatment effect. Given randomization of treatment in the trial, when this model is fit to the trial data, $\gamma_{fa}$ estimates SATE. Note that the coefficient of $FA$ in models with baseline covariates ($X$, $Z$, $V$) that may interact with $A$ or $F$ but not with $FA$ also estimates SATE; such models adjust for covariates when estimating the average treatment effect.

For the sensitivity analyses, we assume a model with effect modification analogous to M1. The full details, which are somewhat more complicated than are informative, are relegated to the S4 Appendix. The key point is that this model includes not only the $FA$ term ($\beta_{fa} F_{ij} A_i$), but also interaction terms of effect modifiers with $FA$ ($\beta_{zfa} Z_i F_{ij} A_i$ and $\beta_{vfa} V_i F_{ij} A_i$). The TATE formula in this case is

$$\mathrm{TATE} = \beta_{fa} + \beta_{zfa}\mathrm{E}[Z|P = 1] + \beta_{vfa}\mathrm{E}[V|P = 1]. \tag{T2}$$

This formula is used with both method 1 and method 2—when the effect modification regression model is fit to the unweighted and weighted trial data, respectively.

While it is usually natural to clarify effect definitions before discussing models used to estimate such effects, in this section we have done the opposite, starting with models first. This choice is intentional because in the current case it is easier to point out the effect definition after explaining the model. As $FA$ is an interaction of treatment arm and time (post- vs. pre-treatment), its coefficient represents a difference in difference, specifically a difference between the two treatment conditions with respect to the difference between post- and pre-treatment outcomes in each condition. This means the individual treatment effect definition here is the effect of treatment on the 'potential outcome change' from before to after treatment. Since pre-treatment outcome is not affected by treatment, this is equivalent to the effect of treatment on the potential post-treatment outcome. That is, the TATE and SATE in this case are exactly the same average effects defined in the previous section.

## Extension 2: Multiplicative effects on potential outcome rate/probability/odds, based on a log/logit link model

In the previous section, we commented that the linear model assumption is not always appropriate. We now extend the methods to cases where log/logit link outcome models are used (e.g., log mean model for a count outcome, log probability or logit model for a binary outcome). Here the individual treatment effect is defined on the multiplicative scale that matches the model (e.g., rate ratio, risk ratio, or odds ratio), and an ATE is defined as the geometric mean of the individual effects. Or equivalently, one could define the individual effect as the corresponding log rate ratio, log risk ratio or log odds ratio, and have the usual definition of ATE as the arithmetic mean of individual effects.

We describe the extension formally here using a binary outcome with logit model (leaving details for the other cases in the S5 Appendix). In this case we assume the logistic causal model

$$\log\left\{\frac{\Pr[Y_i(a)=1]}{\Pr[Y_i(a)=0]}\right\} = \beta_0 + \beta_a a + \beta_x X_i + \beta_z Z_i + \beta_{za} Z_i a + \beta_v V_i + \beta_{va} V_i a, \qquad \text{(M3)}$$

the coefficients of which can be estimated by fitting the logistic regression model

$$\log\left\{\frac{\Pr[Y=1|A,X,Z,V]}{\Pr[Y=0|A,X,Z,V]}\right\} = \beta_0 + \beta_a A + \beta_x X + \beta_z Z + \beta_{za} ZA + \beta_v V + \beta_{va} VA \qquad \text{(R3)}$$

to the original or the weighted trial sample (corresponding to method 1 or 2). Define the individual treatment effect as the odds ratio (OR) of the individual's potential outcomes,

$$\text{TE}_i^{\text{OR}} = \frac{\Pr[Y_i(1)=1]/\Pr[Y_i(1)=0]}{\Pr[Y_i(0)=1]/\Pr[Y_i(0)=0]},$$

or the corresponding log OR

$$\text{TE}_i^{\text{log-OR}} = \log\left(\text{TE}_i^{\text{OR}}\right).$$

Model M3 implies,

$$\text{TE}_i^{\text{log-OR}} = \beta_a + \beta_{za} Z_i + \beta_{va} V_i,$$
$$\text{TE}_i^{\text{OR}} = \exp\left(\beta_a + \beta_{za} Z_i + \beta_{va} V_i\right).$$

On the log OR scale, we have the familiar formula

$$\text{TATE}^{\text{log-OR}} = \beta_a + \beta_{za} \text{E}[Z|P=1] + \beta_{va} \text{E}[V|P=1] \qquad \text{(T3a)}$$

for TATE defined as the average of the individual effects in the target population. Note that 'average' in this definition means arithmetic mean. For effects defined on the OR scale, however, the arithmetic mean does not have a nice formula. Yet we can use another type of average that is natural to quantities on a multiplicative scale and thus is both meaningful and mathematically convenient in this case, the geometric mean. Defining TATE on the OR scale as the geometric mean of the individual ORs in the target population, we obtain

$$\text{TATE}^{\text{OR}} = \exp\left\{\beta_a + \beta_{za} \text{E}[Z|P=1] + \beta_{va} \text{E}[V|P=1]\right\}. \qquad \text{(T3b)}$$

These two TATE formulas serve as the basis for essentially the same sensitivity analyses, as $\text{TATE}^{OR} = \exp(\text{TATE}^{\log\_OR})$.

As an aside, an ATE defined as geometric mean of individual OR effects (termed *average causal OR*) is closely related to the conditional OR routinely estimated by logistic regression with main effects only. Indeed the latter is an approximate estimate of the former (see more about this in the S5 Appendix).

## Illustration

The analyses presented here are merely illustrative. Results should not be taken as clinically informative. In addition to the concise presentation here, the detailed analyses can be found in S6 Appendix, with most of the code (R-code) included in the same appendix, and some Stata code in S7 Appendix. The data are provided in S8 and S9 Appendices.

The trial data include baseline and post-treatment CD4 counts and several baseline characteristics—age in years, sex, race, and severe immune suppression (SIS). The target population data include age groups, sex, and race. See Table 4 for a description of these variables. As noted above, when data include pre- and post-treatment measures of the outcome, there are more than one analysis options. Here we use random intercepts models on the combination of both measures. (For an application modeling post-treatment outcome only, see [6]). The definition of TATE is the average effect of treatment on potential CD4 count gain, and equivalently, the average effect of treatment on potential CD4 count post-treatment, in the target population.

We start by analyzing the trial data. The first question is whether to model CD4 count (and thereby consider its change) on the additive or multiplicative scale. CD4 count is a non-negative variable, so the additive scale may predict out of range, but it is also generally bounded above, which is more restrictive on multiplicative than on additive effects. We follow the HIV literature convention of using the additive scale for CD4 count, noting that our models may be suboptimal in this respect.

The two treatment arms in the trial are similar but the new treatment arm has more female patients and the old treatment arm has a higher SIS proportion (see Table 5). For the moment, assume we have good enough balance; we will come back to this.

SATE (i.e., the average difference, between the two treatments in the trial, in CD4 count change, or equivalently, in post-treatment CD4 counts) is estimated to be 36.6, 95% CI = (28.0,45.2) cells/ml by a simple model with no covariates, and 35.8, 95% CI = (27.3,44.4) by a model that adjusts for baseline covariates. Table 6 shows substantial differences in covariate distribution between the trial sample and target population, suggesting these SATE estimates should not be used as estimates of TATE.

**Table 4. Illustration.** Data availability.

| Variable | Description | Observed in | |
| --- | --- | --- | --- |
| | | trial | population |
| Pre-treatment CD4 count | CD4 count, i.e., number of T-CD4 cells per ml blood, within 10 days of treatment initiation (average if more than one available) | ✓ | |
| Post-treatment CD4 count | CD4 count within 10 days of two months on treatment (average if more than one available) | ✓ | |
| Continuous age | Age in years | ✓ | |
| Categorical age | 4 categories: up to 29, 30-39, 40-49, and 50+ | ✓ | ✓ |
| Sex | Binary, male or female | ✓ | ✓ |
| Race | Dichotomized as White or non-White | ✓ | ✓ |
| Severe immune suppression | Any CD4 count of 50 or lower within the baseline period | ✓ | |

**Table 5. Illustration.** Covariate balance within the trial sample.

|  | New treatment (n = 478) | Old treatment (n = 455) |
|---|---|---|
| Age (mean) | 39.4 | 39.5 |
| Sex (proportion female) | 0.174 | 0.143 |
| Race (proportion nonWhite) | 0.473 | 0.468 |
| SIS (proportion) | 0.437 | 0.475 |

**Table 6. Illustration.** Covariate distribution in the trial sample and target population.

|  | Trial sample (n = 933) | Target population (n = 54,220) |
|---|---|---|
| Age (mean) | 39.5 | not available |
| Age (range) | 16 to 75 | 13 to 80 |
| Age groups (proportions) |  |  |
| 29 and younger | 0.107 | 0.341 |
| 30 to 39 | 0.421 | 0.309 |
| 40 to 49 | 0.348 | 0.247 |
| 50 and older | 0.123 | 0.103 |
| Sex (proportion female) | 0.159 | 0.266 |
| Race (proportion nonWhite) | 0.471 | 0.639 |
| SIS (proportion) | 0.456 | not available |

Through a simple analysis examining interactions of *FA* with baseline covariates (e.g., *FA*\*sex), covariate pairs (e.g., *FA*\*sex\*age), and cross-classifications of categorical covariates (e.g., *FA*\*'nonwhite-female'), we identify that the cross-classification of race and SIS status is an effect modifier (see Table 7). This variation in treatment effects is visualized in Fig 2.

As treatment effects vary across race-by-SIS-status categories, to obtain TATE estimates, we need the proportions of the target population that are in these categories. Race is observed in the target population, with 63.9% being nonwhite. SIS status (in the white and nonwhite groups), however, is not observed. If this were a substantive study, we would comb the literature to seek a plausible range for the proportions with SIS among White and among nonWhite people in the target population. As this is only illustrative, we specify a wide range (0.2, 0.6) for this proportion (which covers the proportion 0.456 in the trial), and assume that this proportion is the same between White and nonWhite people in the target population. With this assumption, this single proportion with SIS in the target population is now our *sensitivity parameter*.

**Table 7. Illustration.** Excerpt from effect modification model[a] fit to trial data.

|  | Estimate | Std. Error | df | t value |
|---|---|---|---|---|
| *FA* | 19.72150 | 7.862622 | 923 | 2.508260 |
| *FA* \* nonWhite-noSIS | 23.44112 | 11.979436 | 923 | 1.956780 |
| *FA* \* White-SIS | 24.93563 | 12.135934 | 923 | 2.054694 |
| *FA* \* nonWhite-SIS | 21.43789 | 11.956149 | 923 | 1.793043 |

[a] The referent category is White-noSIS. The other covariates included are age and sex.

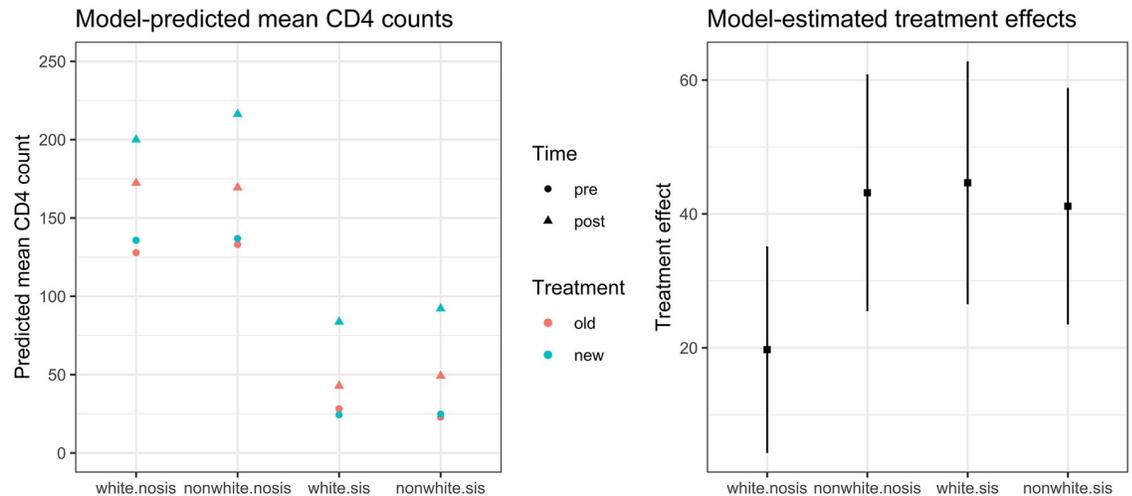

**Fig 2. Illustration.** Visualization of effect modification model.

A comment is warranted on assumption A3, that the range of the treatment effect modifiers in the target population is covered by the trial. In the current example, it is apparent that we do not have a problem with this assumption, because effect modification involves the four race-by-SIS-status categories, all of which are present in the trial sample. What might go unnoticed is that the age range in the trial sample (16 to 75) is smaller than that in the target population (13 to 80), and while we did not detect any treatment effect modification by age in the trial sample, we do not know if treatment effects for those younger than 16 or older than 75 (who are not represented in the trial sample) are the same as or different from treatment effects for those in the trial sample age range; this would be an area to consult clinical experts for their judgment. In the absence of such expert knowledge, were actual age values (rather than just age group) observed in the target population, one option would be to generalize treatment effects only to those aged 16 to 75 in the target population. This target population trimming is not an option in the current case as age values are not observed. In order to move forward, we assume that there is no effect modification by age.

We now apply the *outcome-model-based sensitivity analysis* (method 1) with the specified (0.2, 0.6) range for the sensitivity parameter (proportion with SIS in the target population). This obtains a range of TATE estimates from 36.2, 95% CI = (25.9,46.6) (corresponding to the lower end of the sensitivity parameter range) to 39.3, 95% CI = (30.0,48.6) (corresponding to the upper end of the sensitivity parameter range).

To apply the *weighted-outcome-model-based sensitivity analysis* (method 2), we need to weight the trial data to mimic the target population with respect to covariates observed in both (age group, sex, race). Heeding the recommendation of the two-step procedure, we first adjust the between-arms balance in the trial sample by inverse treatment propensity score weighting. Note that this results in a balance-adjusted SATE estimate of 33.9, 95% CI = (25.4, 42.3), which is very similar to (but slightly smaller than) the original SATE estimates. The effect modification model fit to the balance-adjusted trial data is very similar to that fit to the raw trial data. If we use that model and apply method 1, we obtain TATE estimates ranging from 34.2, 95% CI = (23.5, 44.8) to 37.1, 95% CI = (28.5, 45.7); these are slightly lower than those based on the model fit to the raw trial data. To use method 2, after adjusting within-trial balance, we weight the adjusted trial sample to the target population, using the weighting-by-the-odds procedure described in Table 3. This results in covariate balance shown in Table 8. The ATE in this weighted trial sample (which we will refer to as the $(X, Z)$- adjusted ATE) is 37.4, 95% CI = (27.0, 47.8), which is slightly higher than our SATE estimates. This is consistent with the fact that the weighted trial sample has higher nonWhite and SIS proportions than the original trial sample.

**Table 8. Illustration.** Covariate balance between two trial arms and target population after weighting.

| | Trial: new (n = 478) | Trial: old (n = 455) | Target population (n = 54,220) |
|---|---|---|---|
| Age (mean) | 35.8 | 36.1 | not available |
| Age (range) | 16 to 75 | 16 to 75 | 13 to 80 |
| Age groups (proportions) | | | |
| 29 and younger | 0.349 | 0.333 | 0.341 |
| 30 to 39 | 0.304 | 0.313 | 0.309 |
| 40 to 49 | 0.247 | 0.246 | 0.247 |
| 50 and older | 0.100 | 0.107 | 0.103 |
| Sex (proportion female) | 0.272 | 0.260 | 0.266 |
| Race (proportion nonWhite) | 0.642 | 0.636 | 0.639 |
| SIS (proportion) | 0.515 | 0.522 | not available |

We now fit the same effect modification model to the weighted trial sample (see model fit in Table 9). Interestingly, the coefficients have changed substantially from the model fit to raw data (Table 7). This is not surprising as it could be a result of model misspecification (which we suspected), and it underscores our recommendation to use the weighted method so that the model that is the basis for estimating (sensitivity analysis of) TATE is the model fit to data that is close to the target population with respect to observed variables.

Generally we should not read much into the (lack of) statistical significance of interaction terms in the weighted model because weighting increases variance. However, the underwhelming coefficients in the fitted model above suggest that in the weighted dataset, there is not much differentiation of the group specific effects, and one could argue for using the $(X, Z)$-adjusted ATE as an estimate of TATE.

Or we could choose to proceed with a full implementation of method 2. Based on this effect modification model, we obtain TATE estimates ranging from 38.4, 95% CI = (26.0,50.9) (corresponding to the lower end, 0.20, of the range of the sensitivity parameter) to 37.5, 95% CI = (27.4,47.7) (corresponding to the upper end, 0.60, of the range of the sensitivity parameter).

Results from both methods are plotted in Fig 3. In this particular case, as there is not a clear slope of TATE estimates on the sensitivity parameter, we could combine these results to conclude that when between 20 and 60% of the target population have SIS, we could expect the target population average effect of the new treatment (relative to the old treatment) to be a gain in CD4 count of about 36 to 39 cells per ml (point estimate) with lower and upper confidence bounds of 26 and 51.

**Table 9. Illustration.** Excerpt from effect modification model fit to trial data that have been weighted to the target population.

| | Estimate | Std. Error | z value |
|---|---|---|---|
| *FA* | 32.903852 | 11.93896 | 2.76 |
| *FA* * nonWhite-noSIS | 9.335134 | 15.74757 | 0.59 |
| *FA* * White-SIS | 4.593148 | 14.53369 | 0.32 |
| *FA* * nonWhite-SIS | 3.294465 | 15.46597 | 0.21 |

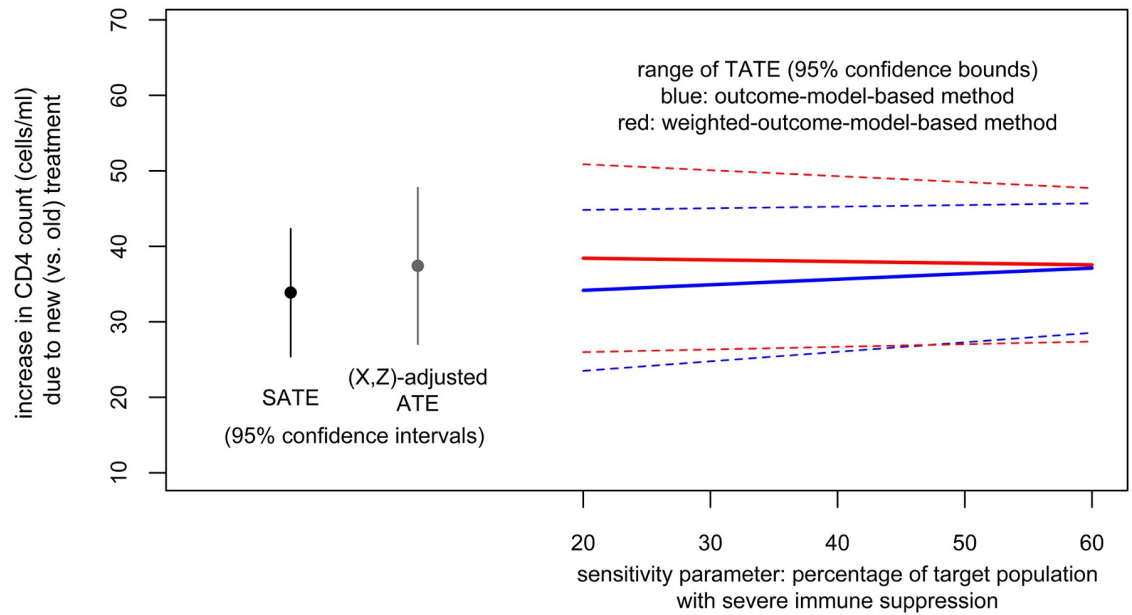

**Fig 3. Illustration.** Results from both methods.

## Discussion

Policy-makers may want to use trial results to inform decisions for specific target populations. There are now methods to calibrate trial results to target populations, increasing potential for evidence-based decision-making [19]. However, these methods generally rely on a strong assumption that all effect modifiers differentially distributed between sample and population are observed and can be adjusted for. In this paper, we describe sensitivity analyses when this is not the case. Sensitivity analyses such as these allow researchers and decision-makers to have more appropriate confidence about population effects.

The methods proposed in this paper address situations where the effect definition matches the model assumed—additive effect with linear model, ratio effect with logit/log link model. While this kind of pairing is natural and commonplace, it is somewhat restrictive; there are situations where the effect definition of policy interest may not be on the same effect scale that best matches the model scientifically deemed appropriate. Our ongoing work aims to provide sensitivity analyses to situations where the model assumed is nonlinear but the effect scale of interest is additive.

That these sensitivity analyses rely on the assumption of a specific causal model deserves discussion, as the assumed model may or may not be close to the truth. Method 2, which combines this model with weighting, provides some protection against model misspecification, and thus has a flavor of double robustness. It is not truly doubly robust, however, because the weighting does not adjust for difference in the distribution of the partially unobserved effect modifier *V*. An area for further development is the search for sensitivity analysis procedures that are more flexible regarding model assumptions.

A challenge in generalization which we commented on is that the effect modifier range coverage assumption (A3) is perhaps commonly not met, because trial samples tend to be less diverse than target populations [20]. Strategies for this situation include combining evidence from multiple trials, combining experimental and non-experimental evidence using cross-design synthesis [21], or redefining the population to the area with overlap [22]. These complex approaches will require adapting sensitivity analyses.

Finally, it is important to note that these sensitivity analyses are not a panacea. It is best to limit the need for them in the first place. This can be done through careful consideration of target populations when designing trials, and efforts to enroll more representative trial samples. If that is not possible, trialists should carefully consider potential effect modifiers and investigate treatment effect heterogeneity. Also, to be able to adjust for effect modifiers when generalizing treatment effects, effect modifiers need to be measured consistently in trials and target population datasets [5, 23]. We encourage trialists to collect covariates in the same way as is done in common population datasets in their fields. When design strategies fall short, the methods discussed here provide a sense for the robustness (or lack thereof) of results when generalizing to target populations.

## Supporting information

**S1 Appendix. Simulations.**
(PDF)

**S2 Appendix. An explanation of why the different weighting procedures apply in the different data scenarios.**
(PDF)

**S3 Appendix. Additional details on the case of effect modifiers not observed in the trial.**
(PDF)

**S4 Appendix. Additional details on extension 1 for random intercepts models.**
(PDF)

**S5 Appendix. Additional details on extension 2 for multiplicative effects and log/logit link models.**
(PDF)

**S6 Appendix. Illustration—Main appendix with explanations and most of the code in R (read this before S7-9).**
(HTML)

**S7 Appendix. Illustration—Stata code for fitting random effects models with probability weights.**
(DO)

**S8 Appendix. Illustration—Trial data.**
(CSV)

**S9 Appendix. Illustration—Target population data.**
(CSV)

## Acknowledgments

This work was supported in part by NSF grant DRL-1335843 (co-PIs E. A. Stuart and R. B. Olsen), PCORI grant ME-150227794 (co-PIs I. Dahabreh and E. A. Stuart), NIDA grant R01-DA036520 (PI R. Mojtabai), and NIAID grant R01AI100654 (PI S. R. Cole). The funders had no role in study design, data collection and analysis, decision to publish, or preparation of the manuscript. The authors acknowledge insightful comments from three referees and the academic editor, which helped improve this work.

## Author Contributions

**Conceptualization:** Trang Quynh Nguyen, Stephen R. Cole, Elizabeth A. Stuart.

**Data curation:** Trang Quynh Nguyen.

**Formal analysis:** Trang Quynh Nguyen.

**Funding acquisition:** Elizabeth A. Stuart.

**Investigation:** Trang Quynh Nguyen.

**Methodology:** Trang Quynh Nguyen, Elizabeth A. Stuart.

**Supervision:** Elizabeth A. Stuart.

**Visualization:** Trang Quynh Nguyen.

**Writing – original draft:** Trang Quynh Nguyen.

**Writing – review & editing:** Trang Quynh Nguyen, Benjamin Ackerman, Ian Schmid, Stephen R. Cole, Elizabeth A. Stuart.